\documentclass[12pt]{article}
\usepackage[pdfstartview=FitH,colorlinks=true,linkcolor=blue,anchorcolor=red,citecolor=magenta,urlcolor=blue]{hyperref}
\usepackage{amsmath,amssymb,titling,authblk}
\usepackage{amscd}
\usepackage{tcolorbox}
\usepackage{amsfonts}
\usepackage{amstext}
\usepackage{amsthm}
\usepackage{bbm}
\usepackage[normalem]{ulem}
\usepackage{appendix}
\usepackage{bbold}
\usepackage{yfonts}
\usepackage{epsfig}
 \usepackage[utf8]{inputenc}
\usepackage[T1]{fontenc}
\usepackage {color}
\usepackage{bbold}
\usepackage{latexsym}
\usepackage{mathrsfs}
\usepackage{soul}
\setstcolor{blue}
\usepackage{cite}
\allowdisplaybreaks[4]
\numberwithin{equation}{section}

\usepackage{bm}
\definecolor{verde}{cmyk}{.83,.21,1,.08}

\newtheorem*{proof*}{Proof}

\advance\hoffset by -1.1truecm
\advance\voffset by -1.0truecm
\newcommand{\be}{\begin{equation}}
\newcommand{\ee}{\end{equation}}
\newcommand{\beqa}{\begin{eqnarray}}
\newcommand{\eeqa}{\end{eqnarray}}
\newcommand{\eqn}[1]{(\ref{#1})}

\numberwithin{equation}{section}

\newcounter{appendice}

\usepackage{lmodern}

\begin{document}


\title{Gravity model from (A)dS Yang-Mills theory}

\author[a,b]{Goffredo Chirco \thanks{goffredo.chirco@unina.it}}
\author[a,b]{Alfonso Lamberti \thanks{alfonso.lamberti@unina.it}}
\author[a,b]{Lucio Vacchiano \thanks{lucio.vacchiano@unina.it}}
\author[a,b]{Patrizia Vitale \thanks{patrizia.vitale@na.infn.it}}

\affil[a]{Dipartimento di Fisica `Ettore Pancini', Universit\`a  di Napoli Federico II, Via Cintia 80126,  Napoli, Italy}
\affil[b]{INFN, Sezione di Napoli, Italy}

\date{}

\maketitle

\vspace{-2cm}
\begin{abstract} 
We investigate the relationship between a one-parameter family of (anti-)de Sitter Yang-Mills models and a model of Einstein-Palatini gravity with matter, realized through Inönü-Wigner contraction of the (A)dS algebra. By setting the group parameter $\alpha$ to zero, the gauge transformation of the potential becomes consistent with the transformation properties of the tetrad form and spin connection. We show that a sector of the Yang-Mills dynamics exists in which the equations decouple. Moreover, a subset of the gauge transformations can be related to diffeomorphisms, leading to the identification of the tetrad field. Finally, the resulting dynamics is consistent with a gravitational dynamics in the first-order formalism.

\bigskip
\textbf{keywords:} {\it (anti-)de Sitter gauge theory,  
first order gravity } 
\end{abstract}
\bigskip
\section{Introduction}
\label{sec1}
The search for a formulation of gravity as a gauge theory has a long history. There is an extensive literature on the subject, starting from the seminal papers  of Utiyama, Sciama, Kibble, Hayashi, and Nakano \cite{Utiyama:1956sy, Sciama:1964wt, Kibble:1961ba, 10.1143/PTP.38.491} and the contributions  by Hehl in the 1980s (see for example the reprint \cite{Hehl:2023khc}), gauging  the Lorentz  and  Poincaré groups. Many of these approaches are reviewed in \cite{Peldan:1993hi}, particularly in the recent textbook \cite{Krasnov:2020lku}. 

The first step toward a gauge-theoretic formulation of gravity typically consists in moving from the second-order metric formulation of the Einstein-Hilbert action to a first-order, or tetradic-Palatini, formulation~\cite{Palatini:1919ffw},
which amounts to choosing new, independent degrees of freedom, tetrads and spin connection, 
and promoting the local Lorentz group to the gauge group. The resulting action does not yield a gauge theory of Yang-Mills type, since the tetrad $e_\mu^a$ is not a gauge connection, while the spin connection $\omega_\mu^{ab}$ is. However, by extending the construction that works in  lower dimensions, where the theory is topological~\cite{Witten:2007kt}, the former can be formulated as a gauge theory of $BF$ type \cite{Birmingham:1991ty}, 
provided the so-called simplicity constraint is implemented for the $B$ field  \cite{Plebanski:1977zz, Krasnov:2010olp}. 

On the other hand, attempts to  formulate gravity as a Yang-Mills  theory for the whole Poincaré group, so to include the tetrads in the gauge connection, encounter a well- known problem: due to the degeneracy of the Cartan-Killing metric of the algebra, the translational sector of the theory is nondynamical, unless the Yang-Mills action is modified or the gauge group is extended (see, for example, \cite{Kawai}, \cite{Aldrovandi:1981mv}). In this framework, many existing proposals in the literature exploit the possibility of obtaining the Poincaré group as a In\"on\"u-Wigner contraction of de Sitter or anti-de Sitter groups \cite{inonu1964group, Inonu}. A related approach is represented by the MacDowell-Mansouri action ~\cite{MacDowell:1977jt} whose geometric interpretation in terms of Cartan geometry is clarified in \cite{Wise_2010}.  For an updated review  see \cite{Krasnov:2020lku}.   

In this paper, we propose yet another formulation of gravity as a one-parameter family of { \it standard} Yang-Mills models on Minkowski space for the (anti) de Sitter group [(A)dS in the following], with the physical content of the theory being recovered in the In\"on\"u-Wigner contraction. 
The choice of pseudo-orthogonal groups offers two obvious advantages: it allows one to encode the would-be geometric fields into components of the extended gauge  connection and it guarantees that the scalar product in the algebra, defined via the Cartan-Killing metric, is invariant under the adjoint action of the group and nondegenerate. 

One main novelty of our proposal is the invariance in the form of the action in the limit $\alpha\rightarrow 0$, being $\alpha$ the contraction parameter from the (A)dS group to the Poincaré group. We shall see that the standard Lagrangian description of the Yang-Mills gauge theory can be applied without modifications, while the geometric interpretation of the gauge fields as  tetrad form and Lorentz connection will be recovered as $\alpha\rightarrow 0$, with the Lorentz subgroup as the residual gauge symmetry. 
The choice of de Sitter or anti-de Sitter group will ultimately be motivated by the fact that they preserve a quadratic form with an appropriately chosen signature, whose four-dimensional reduction is subsequently preserved by the Lorentz group~\cite{Randono:2010cq}.



The idea of using the (A)dS group is  not new. Various works have proceeded along similar lines in the past and the current literature. In~ \cite{PhysRevD.21.1466}, the $SO(3,2)$ gauge symmetry is spontaneously broken  to the Lorentz group, in such a  way as to yield the Einstein-Cartan theory. The Goldstone field associated with the symmetry-breaking mechanism allows the derivation of the gravitational vierbein and spin connection from the original $SO(3,2)$ gauge fields by transitioning to a set of nonlinearly transforming fields through a redefinition that involves the Goldstone field. The work in~\cite{Tseytlin:1981nu} proposes a nonlinear realization of the gauge connection on a fiber bundle, where the fiber is the homogeneous space $SO(1,4)/SO(1,3)$. Then the components of the gauge potential are identified with tetrads and spin connection, leading to Euler-Lagrange equations that have gravitational counterparts in the material sources, linked to curvature and torsion. 

More recently, in \cite{Sobreiro:2011hb}, the Inönü-Wigner contraction, responsible for the transition from the gauge group to the Lorentz group, is related to some mass parameter. Additionally, each gauge field configuration defines an effective geometry through a isomorphism from $\mathbb{R}^4$ to a final manifold $\mathbb{M}^4$, a deformed space. 
Another construction is proposed in \cite{mistretta2023}, where a Yang-Mills type gauge action is analyzed, and the gravitational action emerges as a projection of the five-dimensional space on which the de Sitter group fibration is defined. This projection identifies a natural tetrad, which defines the spacetime volume, thereby breaking the original gauge symmetry to $SO(1,3)$.  Finally, in \cite{Lazzarini:2024ocr}, gravity models are derived from a topological gauge theory which is a combination of topological invariants of a manifold with an underlying Cartan geometry $G/H$, $G$ being the (A)dS group and $H$ the Lorentz subgroup $SO(1,3)$.

Our analysis starts with a standard Yang-Mills gauge theory on a Minkowski base manifold, associated with a one-parameter  family of algebras of pseudo-orthogonal groups, $\mathfrak{g}_{\alpha}$, which recovers the de Sitter, anti-de Sitter, and Poincaré algebras for $\alpha$ being, respectively, positive, negative, or zero. 
In the limit where $\alpha$ tends to zero, the structure group reduces to the Poincaré group but the gauge connection is defined in such a way that the action functional does not change, being independent of $\alpha$. However the interpretation of the gauge fields in the action changes radically. In fact, in the limit, the theory is no longer  a gauge theory for the whole (A)dS group. The local symmetry  of the action reduces to Lorentz making it possible   to interpret the emerging model as a gravitational theory. In this work we  identify and analyze the set of conditions that must be met for such an interpretation to be well grounded. 

The paper is organized as follows.  In Sec.  \ref{sec:YM} we introduce the Yang-Mills action for the one parameter family of Lie algebras. In Sec. \ref{ind D}, we study the differential identities implied by the Bianchi identity of the Yang-Mills action.  
This analysis is needed for ensuring consistency with the Bianchi identities of curvature and torsion of gravity in the limit $\alpha\rightarrow 0$.
In Sec. \ref{Trasf connessione} we analyse the behavior of the different components of the gauge potential under gauge transformations and we show that, in the limit $\alpha\rightarrow 0$, their transformation properties are consistent with   
their interpretation as  tetrad one-form and Lorentz connection.
In Sec. \ref{Eq Lagrange} we derive the Euler Lagrange equations 
in the presence of matter, and we discuss their formal equivalence with the equations of gravity in the first-order formalism.
In Sec. \ref{dalle t ai diff}, we complete the identification of tetrads and Lorentz connection in the limit $\alpha \to 0$, by identifying the diffeomorphisms with a sector of the gauge transformations. 
Finally, we describe the corresponding Euler-Lagrange equations in the limit $\alpha \to 0$, deriving the gravitational equations. 
We conclude by discussing  results and perspectives in Sec. \ref{disc}.

\section{\label{sec:YM}The Yang-Mills action for the (A)dS group}
Let us consider the one-parameter family of Lie algebras $\mathfrak g_{{\alpha}}$\footnote{We indicate with $G_\alpha$ the  family of associated Lie groups. Gauge groups and gauge algebras will be respectively indicated with calligraphic letters  $\mathcal{G}_{\alpha}, g_\alpha$.}, related to  (A)dS  and Poincaré groups, depending on the value of the dimensionless parameter ${\alpha}$.  The  family 
 is characterized by  the Lie brackets 
\begin{align}
     [P_a,P_b]&=\alpha J_{ab}:=\alpha K_{a,b}^{ef}J_{ef} \nonumber
\\
[P_a,J_{bc}]&=\eta_{ab}P_c-\eta_{ac}P_b:=f^e_{a,bc}P_e 
\\
[J_{ab},J_{cd}]&=\eta_{ad} J_{bc} - \eta_{ac} J_{bd} + \eta_{bc} J_{ad} - \eta_{bd} J_{ac}:=f^{ef}_{ab,cd}J_{ef}\nonumber
\end{align}
with generators $ J_{ab}, P_{a}, a,b=0,\dots ,3$ and 
$K^{cd}_{a,b}=\frac{1}{2} (\delta^c_a\delta^d_b-\delta^d_a\delta^c_b)$. This choice of basis singles out the Lorentz subalgebra spanned by the generators $J_{ab}$ (more details in Appendix~\ref{wtr}). 
We  introduce a $\mathfrak g_{{\alpha}}$-valued  connection one-form 
\begin{equation}
\label{connessione Tua}
\Omega=\frac{1}{2}\varpi^{ab}J_{ab}+\sqrt\frac{\lambda}{{{\alpha}}}\vartheta ^a P_{a}\,\,\,\in\Omega^1(\mathcal{M})\otimes\mathfrak{g}_{{\alpha}}\
\end{equation}
with $ \varpi_{ab}$ and $ \vartheta_{a}$ respectively indicating  its  components in the Lie algebra generators $  J_{ab}$ and $P_{a}$. 
The  local expression on the base manifold $(\mathbb{R}^{(1,3)},\eta)$, with $\eta$ the Minkowski metric, is given by $\Omega = \Omega_\mu dx^{\mu}$,  $\mu=0,\dots ,3$. The parameter $\lambda$ has dimensions of  inverse length square and its sign determines the sign of $\alpha$ to ensure that the radicand is positive
\footnote{In addition, notice that as a consequence of the dimensional properties of the coefficient $\lambda$, the two components of the connection have different dimensions: $\varpi^{ab}_\mu$ has dimensions of the inverse of a length, while $\vartheta^a_\mu$ is dimensionless.}.


The curvature two-form $\mathcal{F}$ associated with the connection one-form $\Omega$ 
\begin{equation}
\mathcal{F}=\mathcal{D}_{\Omega}\Omega=d\Omega+\frac{1}{2}[\Omega,\Omega]\,\,\,\,\in\Omega^2(\mathcal{M})\otimes\mathfrak{g}_{\alpha}
\end{equation}
is naturally decomposed into two components, a Lorentz  valued two-form and a second term that is  $P_a$-valued. We have 
\begin{equation}
\label{Curvatura Tua}
\mathcal{F}=\frac{1}{2}(\mathfrak{R}^{ab}+
\lambda\mathfrak{E}^{ab})J_{ab}+\sqrt\frac{\lambda}{{\alpha}}\mathfrak{T}^aP_{a}
\end{equation}
where\begin{equation}
\label{eq C for R}
\mathfrak{R}^{ab}=d\varpi^{ab}+ \frac{1}{4}f^{ab}_{ij,cd}\varpi^{ij}\wedge\varpi^{cd}
=d\varpi^{ab}+\varpi^{a}_{c}\wedge \varpi^{cb}
\end{equation}
is the Lorentz curvature of the connection $\varpi$,  while
\begin{equation}\begin{split}
\label{eq c for T}
\mathfrak{T}^a=d\vartheta^{a}+\frac{1}{2}f^{a}_{cd,b}\varpi^{cd}\wedge\vartheta^{b}=d \vartheta^a +\varpi^{a}_c\wedge\vartheta^c
\end{split}\end{equation}
is the  covariant derivative of $\vartheta$ with respect to $\varpi$ and we have introduced for future convenience the two-form
\begin{equation}
\label{dd}
\mathfrak{E}^{ab}=K^{ab}_{c,d}\vartheta^c\wedge\vartheta^d=\vartheta^a\wedge\vartheta^b.
\end{equation}
 
The Yang-Mills action for the (A)dS group reads
\begin{equation}
\label{Azione GYM}
S_{{YM}}=\frac{\hbar }{2g^2}\int_\mathcal{M}\left( \mathcal{F} \underset{\vcenter{\hbox{\scriptsize,}}}
{\wedge} *\mathcal{F}\right)\,, 
\end{equation}
where the Hodge product is the usual one defined with respect to the Minkowski metric of the base manifold, and the round brackets denote the product in the algebra provided by the Cartan-Killing metric, which is adjoint invariant and nondegenerate for $\alpha \neq 0$. Moreover, the action is globally invariant under Poincaré transformations. 

Starting from \eqref{Curvatura Tua}, we can explicitly expand the Lagrangian density in \eqref{Azione GYM} according to our choice of basis in the algebra (see Appendix~\ref{wtr}). We get a sum of four contributions 
\begin{equation}\begin{split}
\left( \mathcal{F}\underset{\vcenter{\hbox{\scriptsize,}}}\wedge *\mathcal{F}\right)=\frac{1}{4}\mathfrak R^{ab}\wedge *\mathfrak R^{cd}(J_{ab},J_{cd})+\frac{\lambda}{\alpha}\mathfrak T^{a}\wedge *\mathfrak T^{d}(P_{a},P_{d})+\\
+\frac{\lambda}{2} \mathfrak R^{ab}\wedge *\mathfrak E^{cd}(J_{ab},J_{cd})+\frac{\lambda^2}{4}\mathfrak E^{ab}\wedge *\mathfrak E^{cd}(J_{ab},J_{cd}).
\end{split}\end{equation} 
By computing the product in the algebra [see the Appendix in \eqref{Finale (J,J) e (JP)}], the  action finally reads
\begin{multline}
\label{azione finale riscritta}
S_{YM}=\frac{\hbar}{2g^2}\int_\mathcal{M} \frac{1}{2} \mathfrak R^{ab}\wedge *\mathfrak R_{ab}+\lambda \mathfrak T^a\wedge *\mathfrak T_a+\frac{\hbar\lambda}{2g^2}\int_\mathcal{M }\mathfrak R_{ab}\wedge* \mathfrak{E}^{ab}+\frac{\lambda}{2}\mathfrak{E}_{ab}\wedge*\mathfrak{E}^{ab}.
\end{multline}
The first two terms of the action are similar in form to  the Euler classes for the curvature $\mathfrak{R}$ and the two form $\mathfrak{T}$, \footnote{In fact, unlike the Euler classes, which are described using an internal Hodge product, 
these are characterized by a Hodge product  with respect to the  metric of the base manifold.} while the last two are equivalent in form to the Einstein-Palatini action with cosmological constant~\cite{Peldan:1993hi, Palatini:1919ffw}. As for the latter contribution, notice that 
it naturally emerges from  the definition of the gauge connection  \eqref{connessione Tua}. Furthermore, by adding a topological  term to the Yang-Mills action, defined as
\begin{equation}
\label{Holst}
S_{top}=\frac{\kappa}{16\pi^2}\int_\mathcal{M}\left( \mathcal{F}\underset{\vcenter{\hbox{\scriptsize,}}}\wedge \mathcal{F}\right),
\end{equation}
the theory continues to satisfy the fundamental symmetries (gauge invariance and global Poincaré invariance) and would produce boundary effects if the manifold had a boundary. By setting ${\kappa}/{(16 \pi^2)}={\hbar}/{(2\gamma  g^2)}$ (where $\gamma$ will be eventually identified with  the Barbero-Immirzi parameter), we have 
\begin{equation}\begin{split}
\label{dddf}
S_{YM+top}&=\underbrace{\frac{\hbar\lambda}{2g^2}\int_\mathcal M \mathfrak R_{ab}\wedge *(\vartheta^a\wedge\vartheta^b)+\frac{\lambda}{2}\vartheta^a\wedge\vartheta^b\wedge*(\vartheta_a\wedge\vartheta_b)}_{S_{E-P}+{Cosmological\,Constant}}\\&+\underbrace{\frac{\hbar\lambda}{\gamma g^2}\int_\mathcal M \mathfrak R_{ab}\wedge \vartheta^a\wedge\vartheta^b}_{Palatini\,\,Holst\,\, Term}+\underbrace{\frac{\hbar}{2g^2}\int_\mathcal M \frac{1}{2} \mathfrak R_{ab}\wedge *\mathfrak R^{ab}+\lambda\mathfrak T_{a}\wedge *\mathfrak T^{a}}_{{Generalized\,\,Euler- Classes}  }
\\&+\underbrace{\frac{\hbar}{4\gamma g^2}\int_\mathcal M \mathfrak R_{ab}\wedge\mathfrak R^{ab}}_{Pontryagin\,\, Term }+\underbrace{ \frac{\hbar\lambda}{2\gamma g^2}\int_\mathcal M\mathfrak T_{a}\wedge\mathfrak T^{a}-\mathfrak R_{ab}\wedge \vartheta^a\wedge\vartheta^b}_{Nieh-Yan\,\, Term }.
\end{split}\end{equation}
Therefore, the addition  of a topological term to the Yang-Mills action naturally produces further contributions that can be identified with well- known topological invariants of gravity \cite{Corichi:2013zza, Nieh:1981ww, Rezende:2009sv, Montesinos:2001ww}
\footnote{For this reason, with an abuse of notation, we name the topological terms in the same way as their counterparts which will be obtained in the contraction limit.}. 
The question is whether the formal analogy of the (A)dS  Yang-Mills model with first-order Einstein-Palatini gravity can be made physical via a contraction of the structure group to Poincaré, and which conditions should be satisfied for that to happen.
To this aim, the  choice of basis in the (A)dS algebra and its dependence on $\alpha$ play an essential role. In the  limit  $\alpha\rightarrow 0$,  $\mathfrak{g}_\alpha$  reduces to the Poincaré algebra but the  action does not change in form and it is still well defined, being independent of the parameter, despite the fact that 
for $\alpha \to 0$, 
the connection (\ref{connessione Tua}) and the curvature (\ref{Curvatura Tua}) are ill defined. 
Therefore  the action (\ref{azione finale riscritta}) will remain gauge invariant only with respect to the  Lorentz group, suggesting a consistent identification with some formulation of  gravity if the limit is carefully performed.

In the coming sections, we show that the following set of conditions for such identification holds:

\begin{itemize}
\item[i)]
the differential identities implied by the vanishing of the exterior covariant derivative of the curvature two-form, $\mathcal{D}_{\Omega}\mathcal{F}=0$ are consistent with the Bianchi identities of the would-be Riemann curvature and torsion;
\item[ii)] the behavior of the gauge fields  under  the action of the local Lorentz group, in the limit $\alpha\rightarrow 0$, is consistent with their interpretation as tetrad one-form and Lorentz connection;
\item[iii)] a sector of the gauge dynamics is formally consistent with gravitational dynamics in the first-order formalism;
\item[iv)]  the symmetries of gauge theory, in the limit $\alpha\rightarrow 0$, can be identified with those of the geometric fields in the first-order formulation of gravity.
\end{itemize}

\subsection{Differential identities implied by \texorpdfstring{$\mathcal{D}_{\Omega}\mathcal{F}=0$} {}}\label{ind D}

The one-form connection and the two-form curvature introduced in Eqs. \eqref{connessione Tua} and \eqref{Curvatura Tua}, respectively, are local representatives of the Ehresmann connection  and its curvature  on the principal bundle $(P, \mathcal{M}, {G}_\alpha$). The fulfillment of the Cartan structure equations for the curvature  implies the vanishing of the covariant derivative of the associated local two-form, that is
\begin{equation}
\mathcal{D}_{\Omega}\mathcal{F}=d\mathcal{F}+\left[\Omega,\mathcal{F}\right]=0,
\end{equation}
which in turn determines  a set of differential identities, when projected along the Lie algebra generators. In particular, using \eqref{connessione Tua} and \eqref{Curvatura Tua}, we have 
\begin{equation*}\begin{split}
\mathcal{D}_{\Omega}\mathcal{F}&=\frac{1}{2}\left(d\mathfrak R^{ab}+ \frac{1}{2}f^{ab}_{ij,cd}\varpi^{ij}\wedge \mathfrak R^{cd}+\lambda\left( d\mathfrak E^{ab}+\frac{1}{2}f^{ab}_{cd,ef}\varpi^{cd}\wedge \mathfrak E^{ef}+2K^{ab}_{c,d}\vartheta^c\wedge \mathfrak T^d\right)\right)J_{ab}\\
    &+\sqrt\frac{\lambda}{{\alpha}}\left(d\mathfrak T^{a}+\frac{1}{2}f^{a}_{cd,b}\varpi^{cd}\wedge\mathfrak T^{b}+f^{a}_{b,cd}\vartheta ^{b}\wedge\frac{1}{2}\mathfrak R^{cd}+\frac{\lambda}{2}\left( f^a_{b,cd}\vartheta^b\wedge \mathfrak E^{cd} \right)\right)P_{a}.
\end{split}\end{equation*}
Let us analyze the three terms appearing in the covariant derivative. 
We recognize
\begin{equation}\begin{split}
d\mathfrak R^{ab}+ \frac{1}{2}f^{ab}_{ij,cd}\varpi^{ij}\wedge \mathfrak R^{cd}&=\mathcal{D}_{\varpi}\mathfrak R^{ab}
\\
    d\mathfrak E^{ab}+\frac{1}{2}f^{ab}_{cd,ef}\varpi^{cd}\wedge \mathfrak E^{ef}+2K^{ab}_{c,d}\vartheta^c\wedge \mathfrak T^d
   &=\mathfrak T^a\wedge \vartheta^b-\mathfrak T^b\wedge\vartheta^a+\vartheta^a\wedge \mathfrak T^b-\vartheta^b\wedge\mathfrak T^a=0
\\
d\mathfrak T^{a}+\frac{1}{2}f^{a}_{cd,b}\varpi^{cd}\wedge\mathfrak T^{b}+\frac{1}{2}f^{a}_{b,cd}\vartheta^{b}\wedge \mathfrak R^{cd}&=\mathcal{D}_{\varpi}\mathfrak T^{a}-\vartheta^{b}\wedge \mathfrak R^{a}\,_{b}
\end{split}
\end{equation}
and $ f^a_{b,cd}\,\vartheta^b\wedge \mathfrak E^{cd} =0$.  
Therefore,  the Cartan structure equations reduce to
\begin{equation}
\frac{1}{2}\left(\mathcal{D}_{\varpi}\mathfrak R^{ab} \right)J_{ab}+\sqrt\frac{\lambda}{{\alpha}}\left(\mathcal{D}_{\varpi}\mathfrak T^{a}-\vartheta^{b}\wedge \mathfrak R^{a}\,_{b}\right)P_{a}=0, 
\end{equation}
that is, 
\beqa
    \mathcal{D}_{\varpi}\mathfrak R^{ab}&=&0
    \\
    \mathcal{D}_{\varpi}\mathfrak T^{a}-\vartheta^{b}\wedge \mathfrak R^{a}\,_{b}&=&0.
\eeqa
These equations represent the analog of the generalized Bianchi identities for the curvature and the torsion tensor of an affine connection. The identification will become non-formal when the gauge potential acquires a geometric interpretation in terms of  tetrad one-form and the Lorentz connection.

\subsection{\label{Trasf connessione}Transformation properties of connection one-form}\label{sec:4}

To establish the desired interpretation of the gauge connection, it is essential that its components transform correctly under the action of the local Lorentz group in the limit where the full group reduces to the Poincaré group. To check this, let us  first consider gauge transformations of the gauge connection $\Omega$. 
We have
\begin{equation}    \Omega'=g^{-1}\Omega g+g^{-1}dg\, , \qquad \text{with} \quad g\in\mathcal{G}_{\alpha}
\end{equation}
whose infinitesimal form reads
\begin{equation}\begin{split}
\Omega'=\Omega+\varepsilon^{ab}[\Omega,J_{ab}]+\varepsilon^a[\Omega,P_a]+d\varepsilon^{ab}J_{ab}+d\varepsilon^aP_a.
\end{split}\end{equation}
By setting
\begin{equation}
\Omega' = \frac{1}{2}\varpi'^{ab}J_{ab}+\sqrt{\frac{\lambda}{\alpha}}\vartheta'^a P_{a}
\end{equation}
we find 
\begin{equation}\label{trasf d g iniziale}
\begin{split}
\frac{1}{2}\delta\varpi^{ef}&=f_{ab,cd}^{ef}\frac{1}{2}\varpi^{ab}\varepsilon^{cd}+\sqrt{\alpha\lambda}K^{ef}_{a,d}\vartheta^a\varepsilon^d+d\varepsilon^{ef}
\\
\delta\vartheta ^e&=f_{a,cd}^{e}\vartheta^a\varepsilon^{cd}+\sqrt{\frac{\alpha}{\lambda}}(f^e_{ab,d}\frac{1}{2}\varpi^{ab}\varepsilon^d+d\varepsilon^e)
\end{split}\end{equation}
where $\delta\varpi^{ab}=\varpi'^{ab}-\varpi^{ab}$ and $\delta\vartheta^a=\vartheta'^a-\vartheta^a$.
Let us now take the limit for $\alpha \rightarrow 0$. In this limit, $G_\alpha$ reduces to the Poincaré group $SO(1,3)\ltimes T_4$, and Eqs.  \eqref{trasf d g iniziale} become 
\begin{equation}\begin{split}
\label{Trasf della connessioone GYM}
        \frac{1}{2}\delta\varpi^{ef}&=d\varepsilon^{ef}+f_{ab,cd}^{ef}\frac{1}{2}\varpi^{ab}\varepsilon^{cd}
        \\
        \delta\vartheta ^e &=f_{a,cd}^{e}\vartheta^a\varepsilon^{cd}.
\end{split}\end{equation}
Hence, the component $\varpi$ of the connection transforms nonhomogeneously under Lorentz transformations, just like the Lorentz connection, while the component $\vartheta$  transforms like a Lorentz vector, similarly to  tetrad one-forms. In other words, in the limit $\alpha \rightarrow 0$,  $\vartheta$ is no longer a gauge connection, whereas $\varpi$ is, and the residual gauge group is the Lorentz group.

\section{Euler-Lagrange equations of YM theory\label{Eq Lagrange}}
The transformation law \eqref{Trasf della connessioone GYM} is not sufficient to establish a complete identification of the gauge fields  as the geometric fields of first-order gravity. It is necessary to understand whether the Euler-Lagrange equations of the Yang-Mills theory, in some sector, are compatible with the gravitational dynamics. As the next step, we analyze the Yang-Mills equations including source terms. 

Assuming that the base manifold has no boundary, the functional derivative of the action \eqref{Azione GYM} with respect to the connection \eqref{connessione Tua} is given by
\begin{equation}
\label{Va YM}
\frac{\delta }{\delta \Omega}  S_{YM}= \frac{\hbar }{2g^2} \mathcal{D}_{\Omega} * \mathcal{F}
\end{equation}
which, in components, gives 
\beqa
\label{A3}
   \frac{\delta}{\delta{\vartheta}_a}S_{YM}&=&\frac{\lambda\hbar}{2g^2}\left[\mathcal{D}_\varpi(*\mathfrak T^a)+\vartheta_b\wedge*\mathfrak R^{ab}+\lambda\vartheta_b\wedge*\mathfrak E^{ab}\right]
\\
\label{A4}
\frac{\delta}{\delta\varpi_{ab}}S_{YM}&=&\frac{\hbar}{4g^2}\left[\mathcal{D}_\varpi(*\mathfrak R^{ab})+\lambda \mathcal{D}_{\varpi}(* \mathfrak{E}^{ab})+\lambda(\vartheta^a\wedge*\mathfrak T^b-\vartheta^b\wedge*\mathfrak T^a)\right].
\eeqa
We observe that the dynamics of the curvature $\mathfrak R$ in \eqref{A3} is characterized by a term related to the covariant derivative of the two-form $\mathfrak T$, while Eq \eqref{A4} for the two-form $\mathfrak T$ is characterized by the presence of the derivatives of the curvature $\mathfrak R$, as well as by a linear term in $\mathfrak T$. These terms arise directly from the presence of the Euler classes in the action \eqref{azione finale riscritta}.

To include sources in the Yang-Mills equations, given a gauge-invariant source action, we define the three-forms as
\begin{equation} \label{tre forma tau} \Theta^a = \frac{\delta S_{sources}}{\delta \vartheta_a},  \end{equation}
\begin{equation} \label{tre forma spin} {\Sigma}^{ab} = \frac{\delta S_{sources}}{\delta \varpi_{ab}} \end{equation}
that can be described in a local basis as
\begin{equation}   \label{Te} \Theta^a=\Theta^a_{[\sigma\rho\lambda]}dx^\sigma\wedge dx^\rho\wedge dx^\lambda=\frac{1}{3!}\mathscr{T}^a_\mu\varepsilon_{\sigma\rho\lambda}^{\mu}dx^\sigma\wedge dx^\rho\wedge dx^\lambda
\end{equation}
\begin{equation}
    \label{Se}\Sigma^{ab}=\Sigma^{ab}_{[\sigma\rho\lambda]}dx^\sigma\wedge dx^\rho\wedge dx^\lambda=\frac{1}{3!}\mathcal S ^{ab}_\mu\varepsilon_{\sigma\rho\lambda}^{\mu}dx^\sigma\wedge dx^\rho\wedge dx^\lambda
\end{equation}
where $\mathscr{T}^a_\mu$ and $\mathcal{S}^{ab}_\mu$ are the energy-momentum and spin tensors. 

Accordingly, the variation of the source action with respect to the entire connection can be described by the current three-form
\begin{equation}
  J=\frac{1}{2}\Sigma^{ab} J_{ab}+\frac{1}{\sqrt{\lambda\alpha}}\Theta^aP_a
\end{equation}
in such a way that its components correspond to the source three-forms defined by \eqref{tre forma tau} and \eqref{tre forma spin}. Therefore, we can write
\begin{equation}
    \delta S_{source}=\int_\mathcal{M}\left( \delta\Omega \underset{\vcenter{\hbox{\scriptsize,}}}
{\wedge} J\right).
\end{equation}
Then, given the total action $ S_{T}=S_{YM}+S_{sources}$, the variation 
\begin{equation}
\label{Va T}
    \delta S_T=\frac{\hbar }{2g^2}\int_\mathcal{M}\left( \delta\Omega \underset{\vcenter{\hbox{\scriptsize,}}}
{\wedge} \mathcal{D}_{\Omega}*\mathcal{F}\right)\,+\int_\mathcal{M}\left( \delta\Omega \underset{\vcenter{\hbox{\scriptsize,}}}
{\wedge}  J\right)
\end{equation}
leads to the equation of motion
\beqa 
\label{Py}
\frac{\hbar}{2g^2}\,\mathcal{D}_{\Omega}*\mathcal{F}=- J,
\eeqa
which, in components, reads
  \beqa \label{ Eq 1}
   \mathcal{D}_\varpi(*\mathfrak T^a)+\vartheta_b\wedge*\mathfrak R^{ab}+\lambda\vartheta_b\wedge*\mathfrak E^{ab}&=&\chi\Theta^a\,,
\\
\label{Eq 2}
\mathcal{D}_\varpi(*\mathfrak R^{ab})+\lambda \mathcal{D}_{\varpi}(* \mathfrak{E}^{ab})+\lambda(\vartheta^a\wedge*\mathfrak T^b-\vartheta^b\wedge*\mathfrak T^a)&=&\lambda\chi{\Sigma}^{ab}\,,
\eeqa
where $ \chi=-{2g^2}/{\hbar\lambda}$.


In addition to Eqs. \eqref{ Eq 1}, \eqref{Eq 2}, there is another set of equations directly provided by the gauge theory.
Specifically, as a consequence of \eqref{Py},  the  three-form current is conserved, yielding
\begin{equation}
\begin{aligned}
0=    \mathcal D_{\Omega} J&= dJ+[\Omega,J]=\left(\frac{1}{2}d\Sigma^{ab}+\frac{1}{4}f^{ab}_{cd,ef}\varpi^{cd}\wedge\Sigma^{ef}+\sqrt{\alpha\lambda}K^{ab}_{c,d}\vartheta^c\wedge\frac{1}{\sqrt{\alpha\lambda}}\Theta^d \right)J_{ab}\\&
        +\left(\frac{1}{\sqrt{\alpha\lambda}}\left[d\Theta^{a}+\frac{1}{2}f^{a}_{cd,b}\varpi^{cd}\wedge\Theta^b\right]+\frac{1}{2}\sqrt{\frac{\lambda}{\alpha}}f^a_{c, bd}\vartheta^c\wedge\Sigma^{bd}\right)P_a .
\end{aligned}
\end{equation}
Thereby, we get the conservation equations
\beqa
\label{e}
   \mathcal{D_{\varpi}}\Sigma^{ab}+\vartheta^a\wedge \Theta^b-\vartheta^b\wedge \Theta ^a&=&0,
    \\
 \label{f}   \mathcal{D}_{\varpi}\Theta^a+\lambda\vartheta_b\wedge {\Sigma}^{ab}&=&0.
\eeqa


Equations. \eqref{e} and \eqref{f}
are similar in form to the conservation laws obtained in the Einstein-Cartan-Sciama-Kibble (ECSK) approach~\cite{Krasnov:2020lku,  PhysRevD.21.3269}. However,
in our case they stem from the gauge conservation of the three-form $J$ imposed by Yang-Mills theory.


At this stage, 
the analogy of Eqs. \eqref{ Eq 1} and \eqref{Eq 2} with the Einstein-Palatini equations becomes apparent in form. However, unlike the Palatini formulation, the algebraic relation of $\mathfrak T$ with spin is not a direct consequence of the equations of motion. Nevertheless, we can show that there exists a sector of the dynamics, identified by a Lorentz gauge-covariant condition, with respect to which this relation is reproduced.

Indeed, by assuming
\be \label{dynchoice}
\mathcal{D}_\varpi*\mathcal{F}^{ab}=\mathcal{D}_\varpi(*\mathfrak R^{ab})+\lambda \mathcal{D}_{\varpi}(* \mathfrak{E}^{ab})=0,
\ee
Eqs. \eqref{ Eq 1}-and \eqref{Eq 2} result in
\beqa
\label{t3}
\mathcal{D}_\varpi(*\mathfrak T^a)+\vartheta_b\wedge*\mathfrak R^{ab}+\lambda\vartheta_b\wedge*\mathfrak E^{ab}&=&\chi\Theta^a,
\\
\label{t4}
\vartheta^a\wedge*\mathfrak T^b-\vartheta^b\wedge*\mathfrak T^a&=&\chi{\Sigma}^{ab}.
\eeqa
In particular, in this case, the equations of motion describe a nonpropagating torsion field $\mathfrak T$, due to the algebraic relationship between torsion and spin that is consequently established.

By inspecting Eqs.~\eqref{ Eq 1} and \eqref{Eq 2} it is conceivable that different choices besides $\mathcal{D}_\varpi*\mathcal{F}^{ab}=0$ can lead to a consistent reduction of the dynamics compatible with gravity. We shall investigate these alternatives in future work.

\section{Geometric character of the gauge fields for \texorpdfstring{$\alpha \to 0$}{}}\label{dalle t ai diff}
So far, we have considered a standard Yang-Mills theory on Minkowski spacetime with structure group 
${G}_\alpha$ 
and a Yang-Mills action given by   \eqref{Azione GYM}, that is invariant under global Poincaré transformations.  
The connection one-form $\Omega$ decomposes along the Lie algebra generators in terms of 
$\vartheta$ and $ \varpi$, which transform appropriately under the action of the gauge group. Moreover they behave as covariant vectors under global Poincaré transformations. 

In Sec. \ref{sec:4}, we remarked that the $P$-gauge translations no longer encode the gauge transformations of $\vartheta$ in the limit $\alpha \to 0$ [see Eqn.~\eqref{Trasf della connessioone GYM}]. We are going to show that the inhomogeneous part of the $P$-gauge transformations can be reinterpreted in terms of diffeomorphisms. 

\subsection{From gauge translations to diffeomorphisms: Soldering}
Let us start by recalling the expression for the gauge transformations of the connection components under the action of the gauge group $\mathcal{G}_\alpha$. We have
\beqa
   \delta \sqrt{\frac{\lambda}{\alpha}}\vartheta^a_\mu&=&\partial_\mu \varepsilon^a+ \varepsilon^c\frac{1}{2}\varpi^{de}_\mu f^a_{de,c}+\sqrt{\frac{\lambda}{\alpha}}\varepsilon^{de}\vartheta^{c}_\mu f^a_{c,de}\nonumber\\
   &:=&\delta_{P(\varepsilon^c)}(\vartheta_\mu^a)+\sqrt{\frac{\lambda}{\alpha}}\delta_{J(\varepsilon^{cd})}(\vartheta_\mu^a),\label{1}\\
\frac{1}{2}\delta \varpi^{ab}_\mu&=&\partial_\mu \varepsilon^{ab}+\varepsilon^{cd}\frac{1}{2}\varpi^{ef}_\mu f^{ab}_{ef,cd}+\sqrt{\alpha\lambda} K^{ab}_{d,c}\vartheta_\mu^d\varepsilon^c\nonumber\\
&:=&\delta_{J(\varepsilon^{cd})}(\frac{1}{2}\varpi_\mu^{ab})+\sqrt{\alpha\lambda} K^{ab}_{d,c}\vartheta_\mu^d\varepsilon^c, \label{2}
        \eeqa
where we denoted by $\varepsilon^a$ and $\varepsilon^{ab}$ the coordinate-dependent parameters of the gauge transformations\footnote{Hereafter, $\delta$ without subscripts (GCT) or (CGCT) will always refer to gauge transformations, and their parameters, functions of spacetime coordinates, will be enclosed in parentheses.} and we have singled out the different contributions along the generators of the Lie algebra. 
At the same time, the connection transforms infinitesimally under general coordinate transformation (GCT) $y^\mu(x)=x^\mu -\xi^\mu(x)$, according to  
\begin{equation}
\delta_{GCT(\xi)}\Omega^I_\mu=\mathcal{L}_\xi\Omega^I_\mu=\xi^\nu\partial_\nu \Omega_\mu^I+ \Omega_\nu^I\partial_\mu\xi^\nu.\end{equation}
Therefore, for each connection component, we have that
\be
\begin{split}
\label{dIFFEO}
\delta_{GCT(\xi)}\vartheta^a_\mu&=\xi^\nu\partial_\nu \vartheta_\mu^a+ \vartheta_\nu^a\partial_\mu\xi^\nu,
\\
\delta_{GCT(\xi)}\varpi^{ab}_\mu&=\xi^\nu\partial_\nu \varpi_\mu^{ab}+ \varpi_\nu^{ab}\partial_\mu\xi^\nu.
\end{split}
\ee
In order to study the interplay of gauge and coordinate transformations, it is convenient to consider a \emph{covariant} generalization of the latter (CGCT) 
\cite{Freedman_VanProeyen_2012}. For the full gauge potential this is defined as
\begin{equation}
\label{CGCT}
\delta_{CGCT(\xi)}\Omega^I_\mu=\delta_{GCT(\xi)}\Omega^I_\mu-\delta_{(\xi^\rho\Omega_{\rho})}\Omega^I_\mu\, 
\end{equation}
and consists of the difference between a diffeomorphism and a gauge transformation with gauge parameter $\xi_\rho \Omega_\rho$. 
This definition  takes care of the unwanted derivatives of the parameters under coordinate transformations. Moreover, it provides a convenient rewriting of the GCT in terms of components of the curvature two-form \eqref{Curvatura Tua}. 

Expanding the definition in~\eqref{CGCT} for the components of $\Omega$, we obtain
\be\begin{split}
\label{5}
    \delta_{CGCT(\xi)} \sqrt{\frac{\lambda}{\alpha}} \vartheta^a_\mu &= 
    \delta_{UGT(\xi)} \sqrt{\frac{\lambda}{\alpha}} \vartheta^a_\mu \\
    &- \left( \partial_\mu ( \xi^\rho \sqrt{\frac{\lambda}{\alpha}} \vartheta^a_\rho )
    + ( \xi^\rho \sqrt{\frac{\lambda}{\alpha}} \vartheta^c_\rho ) 
    \frac{1}{2} \varpi^{de}_\mu f^a_{de,c} 
    + ( \xi^\rho \frac{1}{2} \varpi^{de}_\rho) 
    \sqrt{\frac{\lambda}{\alpha}} \vartheta^c_\mu f^a_{c,de} \right)
\end{split}
\ee
and
\begin{equation}\begin{split}
\label{6}
        \delta_{CGCT(\xi)}(\frac{1}{2}\varpi^{ab}_\mu)&=\delta_{GCT(\xi)}(\frac{1}{2}\varpi^{ab}_\mu)\\
        &- \left(\partial_\mu(\xi^\rho \frac{1}{2}\varpi^{ab}_\rho )+ (\xi^\rho \frac{1}{2}\varpi^{cd}_\rho )\frac{1}{2}\varpi^{ef}_\mu f^{ab}_{ef,cd}+(\xi^\rho \vartheta^c_\rho)\vartheta^{d}_\mu \lambda K^{ab}_{d,c}\right),            \end{split}\end{equation} 
which, on using (\ref{1}) and (\ref{2}), simplify to
\be
\begin{split}
        \label{Covariant CGt}
        \delta_{CGCT(\xi)}\vartheta^a_\mu&=\delta_{GCT(\xi)}\vartheta^a_\mu-\delta_{P(\xi^\rho\vartheta^c_\rho)}(\vartheta_\mu^a)-\delta_{J(\xi^\rho\varpi^{cd}_\rho)}(\vartheta_\mu^a),
        \\
        \delta_{CGCT(\xi)}\varpi^{ab}_\mu&=\delta_{GCT(\xi)}\varpi^{ab}_\mu-\delta_{J(\xi^\rho\frac{1}{2}\varpi^{cd}_\rho)}(\varpi_\mu^{ab})-2\lambda\xi^\rho \vartheta^c_\rho\vartheta_\mu^{d}K_{d,c}^{ab}.   
    \end{split}
\ee
On the other hand,  by replacing the first of Eqs.  (\ref{dIFFEO}) in (\ref{5}), we get
\begin{equation}
\begin{split}
\label{Ab}\delta_{CGCT(\xi)}\vartheta^a_\mu=\xi^\rho\mathfrak{T}^a_{\rho\mu},
\end{split}
\end{equation}
and by replacing the second of Eqs.  (\ref{dIFFEO}) in  (\ref{6}), we have 
\begin{equation}\begin{split}
\label{B} \delta_{CGCT(\xi)}\varpi^{ab}_\mu=\xi^\rho\mathfrak R^{ab}_{\rho\mu}-2\lambda\xi^\rho \vartheta^c_\rho\vartheta_\mu^{d}K_{d,c}^{ab}.\end{split}\end{equation}
Therefore, by comparing Eqs. \eqn{Covariant CGt} with Eqs. \eqn{Ab}, \eqn{B}, we obtain
\begin{equation}
\label{diFFEO2}
\begin{split}
\delta_{GCT(\xi)}\vartheta^a_\mu-\delta_{P(\xi^\rho\vartheta^c_\rho)}(\vartheta_\mu^a)-\delta_{J(\xi^\rho\varpi^{cd}_\rho)}(\vartheta_\mu^a)&=\xi^\rho\mathfrak{T}^a_{\rho\mu},
\\
\delta_{GCT(\xi)}\varpi^{ab}_\mu-\delta_{J(\xi^\rho\frac{1}{2}\varpi^{cd}_\rho)}(\varpi_\mu^{ab})&=\xi^\rho\mathfrak R^{ab}_{\rho\mu}.
\end{split}
\end{equation}
As anticipated, CGCT allow one to rewrite the diffeomorphisms in terms of gauge transformations and covariant tensors. The first of Eqs. \eqn{diFFEO2}, appropriately rearranged, gives the  {\it soldering equation}
\begin{equation}\begin{split}
\label{EQ di saldatura}\delta_{P(\xi^\rho\vartheta^c_\rho)}(\vartheta_\mu^a)= \delta_{GCT(\xi)}\vartheta^a_\mu-\xi^\rho\mathfrak{T}^a_{\rho\mu}-\delta_{J(\xi^\rho\varpi^{cd}_\rho)}(\vartheta_\mu^a)
\end{split}\end{equation}
which explicitly shows that the difference between a $P$-gauge translation with parameter  $\varepsilon^c=\xi^\rho\,\vartheta^c_\rho$ and a diffeomorphism consists of a curvature term and a Lorentz gauge transformation. 
In the limit $\alpha \to 0$, the relation between gauge and diffeomorphism parameters, together with the transformation law  (\ref{Trasf della connessioone GYM}), allows interpreting the $P$-gauge fields $\vartheta^a_\mu$ as tetrads. So for $\alpha\rightarrow 0$, the following  identifications hold:
\begin{equation}
\label{id campi}
\begin{cases}
\vartheta^a\rightarrow \theta^a 
\,\,\,
\text{tetrad one-form}
\\
\varpi^{ab}\rightarrow \omega^{ab}
\,\,\,
\text{Lorentz connection one-form}
\end{cases}
\end{equation}
Consequently, by recalling the definition of the fields in Eqs. \eqref{eq C for R} and \eqref{eq c for T}, we can give a consistent geometric interpretation to the components of the curvature two-form $\mathcal F$:
\begin{equation}
\label{id campi 1}
\begin{cases}
\mathfrak{T}^{a}\rightarrow T^{a}
\,\,\,
\text{
torsion two-form }
\\
\mathfrak{R}^{ab}\rightarrow R^{ab}
\,\,\,
\text{Lorentz curvature two-form }
\end{cases}
\end{equation}

\subsection{\label{Equazioni di Einstein} Recovering  gravitational dynamics}

We are now able to show that the Euler-Lagrange equations of the Yang-Mills action in~\eqref{Azione GYM}, reduced to ~\eqref{t3}, \eqref{t4}, are consistent with gravitational equations  once the gauge fields are identified as described in \eqref{id campi}. 
Equations \eqref{t3} and\eqref{t4}, in components, read
\begin{align}
\label{Eqf2}
\left(\frac{1}{2} \mathcal{D}_{\sigma}^\varpi \tilde{\mathfrak T}^a_{\rho\lambda}+\frac{1}{4}\mathfrak R_{[\mu\nu]}^{ab}\vartheta_{b\sigma}\varepsilon^{\mu\nu}_{\rho\lambda}+  \frac{\lambda}{2}\vartheta_{b\sigma}\vartheta^a_{\mu} \vartheta^b_\nu\varepsilon^{\mu\nu}_{\rho\lambda}\right)\varepsilon^{\alpha\sigma\rho\lambda} &=\chi\Theta^a_{[\sigma\rho\lambda]}\varepsilon^{\alpha\sigma\rho\lambda},
\\
\label{Eqf22}
\tilde{\mathfrak{T}}^b_{\rho\lambda}\vartheta^a_\sigma\varepsilon^{\alpha\sigma\rho\lambda}&=\chi\Sigma^{ab}_{[\sigma\rho\lambda]}\varepsilon^{\alpha\sigma\rho\lambda},
\end{align}
 where we defined $\tilde{\mathfrak T}^a_{\rho\lambda}=\frac{ 1}{2} \varepsilon^{\mu\nu}_{\rho\lambda}\mathfrak T^a_{[\mu\nu]}$ and we used the explicit expressions $\mathfrak R^{ab}_{[\mu\nu]}=\partial_\mu\varpi^{ab}_{\nu}-\partial_\nu\varpi^{ab}_\mu+\varpi^{a}_{\mu\,c}\varpi^{cb}_{\nu }-\varpi^{a}_{\nu\,c}\varpi^{cb}_{\mu}$,  $\mathfrak T^a_{[\mu\nu]}=\partial_\mu\vartheta^{a}_{\nu}-\partial_\nu\vartheta^{a}_\mu+\varpi^{a}_{\mu\,c}\vartheta^{c}_{\nu }-\varpi^{a}_{\nu\,c}\vartheta^{c}_{\mu}$.
 Now, using the definitions of the sources  given in \eqref{Te} and \eqref{Se}, in the limit $\alpha \to 0$ where the fields acquire a geometric interpretation, the Eqs.  \eqref{Eqf2} and \eqref{Eqf22} can be written as
\begin{align}
\label{Eqf3}
\mathcal{D}_\mu^{\omega}\, T_{[\beta\nu]}^a \eta^{\mu\nu}+ R^{ab}_{[\beta\nu]}\theta_{b\mu}\eta^{\mu\nu}+\lambda(\theta_{b\mu}\theta^a_\beta \theta^b_\nu-\theta_{b\mu}\theta^a_\nu\theta^b_\beta)\eta^{\mu\nu}=\,&\chi \mathscr{T}^a_\beta ,  
\\
\label{Eqf33}
2\theta^a_\mu T^b_{[\beta\nu]}\eta^{\mu\nu}=\,&\chi \mathcal{S}^{ab}_{\beta}.
\end{align}
In the same limit, the geometric field $\theta^a_\mu$ and its dual $ \theta^{a\mu}=\eta^{\mu\nu}\theta^a_\nu$ can be used to define a new metric field 
\begin{equation}
g^{\mu\nu}=\eta_{ab}\theta^{a\mu}\theta^{b\nu}\, ,
\end{equation}
with inverse $g_{\mu\nu}=[g^{\mu\nu}]^{-1}$. Now the dual of $\theta^{a\mu}$ with respect to the new metric field is given by  
\begin{equation}    \overline{\theta}^a_\mu=g_{\mu\nu}\theta^{a\nu}
\end{equation}
with ${\theta}^\mu_a,\overline{\theta}^a_\mu$, by construction, satisfying the tetrad relations
\begin{equation}
    \overline\theta^a_\mu\, \theta^\nu_a =\delta^\nu_\mu\,\, , \quad     \overline\theta^a_\mu\, \theta^\mu_b =\delta^a_b.
\end{equation}
In terms of the new fields, Eqs.~\eqref{Eqf3} and\eqref{Eqf33} take the form
\begin{multline}
\label{eq4}
R^a_\beta-\frac{1}{2} \overline \theta^a_\beta R+\Lambda^\rho_\beta \overline{\theta}^a_\rho=\chi(\mathscr{T}^a_\beta-\frac{1}{2}\overline{\theta}^a_\beta\mathscr{T})-\mathcal{D}_{\omega}^\nu T^a_{[\beta\nu]}+\frac{1}{2}\overline\theta^a_\beta(\theta^\lambda_c \mathcal{D}_{\omega}^\sigma T^c_{[\lambda\sigma]}),
\end{multline}
\begin{equation}
\label{eq44} 
T^b_{[\beta\nu]}=\frac{\chi}{2} \, \overline\theta_{a\nu} \mathcal{S}^{ab}_\beta,
\end{equation}
with
\begin{equation}
\Lambda^\rho_\beta=\lambda(gg^\rho_\beta-g^\alpha_\beta g^\rho_\alpha)-\frac{1}{2}\lambda(g^2-g^\alpha_\lambda g^\lambda_\alpha)\delta^\rho_\beta,
\end{equation}
and we use $g=\eta_{\mu\nu}g^{\mu\nu}$ and $g^\alpha_\mu=\eta_{\mu\beta} g^{\beta\alpha}$. Moreover,  $R^{a}_{\beta}=\theta^\nu_b R^{ab}_{[\beta\nu]}$ is the Ricci tensor, $R=\theta^\beta_a\theta^\nu_b R^{ab}_{[\beta\nu]}$ is the scalar curvature, while $\mathscr{T}=\theta^\beta_a\mathscr{T}^a_\beta$ is the trace of the energy-momentum tensor. This shows that, in the case where torsion is algebraically related to spin, the equations of motion for the curvature exhibit an additional source term in the energy-momentum tensor, which depends on the spin tensor. Moreover, the cosmological term becomes nontrivial, including contributions from self-interactions of the induced metric field and coupling terms with the Minkowski background metric.

\section{Discussion}\label{disc}

One of the primary motivations for investigating gauge formulations of gravity is searching for a viable candidate for a quantum theory. In this sense, Yang-Mills theories are a natural choice for their renormalization properties and their success in describing all other fundamental interactions. The question is how much geometrical and physical structure a Yang-Mills theory shares with gravity.
This work investigates the problem by analyzing the inherent geometric structures shared by Yang-Mills models and gravity in its first-order formulation.


As for the parameters of the theory, in the limit $\alpha\rightarrow 0$ we have 
$  \chi=-{2g^2}/{\hbar\lambda}={16\pi G_N}/{c^3}$. It follows that $\lambda$ and $g^2$ can be expressed in relation to one another through the squared Planck length $l^2_p=\hbar G_N/c^3$. As a consequence of the relation between $g^2$ and $\lambda$ and the convention adopted in the construction of the action, $\lambda$ must be negative, which in turn constrains $\alpha$ to be negative, as shown in Eq \eqref{connessione Tua}. This selection leads to $SO(2,3)$ as the family gauge structure group.
However, adopting a different sign convention for the action would have led to a positive $\alpha$, thereby selecting the $SO(1,4)$ family. This highlights the fact that the sign of $\alpha$ does not play an essential role in the construction, which can be developed in a fully general manner for both the de Sitter and anti-de Sitter cases. 

A few more comments are in order. As discussed in \cite{PhysRevLett.80.4851}, the reliance of modern physics on two closely related but separate principles, i.e. general covariance and gauge invariance, may be unsatisfactory. Unifying these principles into a single framework is a task, and two main approaches emerge: one views gauge symmetry as a manifestation of general covariance in higher dimensions, while the other interprets general covariance as a form of gauge symmetry. The results presented in this paper go in the direction of the second approach. This occurs in the limit $\alpha\rightarrow 0$, where the gauge nature of the theory changes, providing a geometric interpretation of gauge potential components as the tetrad and the Lorentz connection.

Since the model is obtained from a standard Yang-Mills action on Minkowski spacetime, well-established quantization techniques of non-Abelian gauge theories apply, although the issues related to the noncompactness of the structure group remain open, { in terms of  both unitarity and renormalizability of the quantum theory \cite{Tseytlin:1981nu}. The results for the gravity sector of the theory should be analized while performing the limit $\alpha\rightarrow 0$.

The possible physical meaning of the proposed mechanism, which reduces the gauge dynamics to the gravitational sector, deserves further investigation. One possibility could be a Higgs-like mechanism, which would involve a spontaneous breaking of the gauge symmetry to the residual Lorentz gauge symmetry, independently of the $\alpha$ limit (see e.g. \cite{Ivanenko:1983fts, Percacci:1984bq}). Another possibility concerns the idea that the gauge parameters of the theory, $\lambda$ and $g$, be renormalized during quantization and would exhibit a running dependence on $\alpha$. The latter,  as a dimensionless parameter, could be linked to the energy scale only if the Planck length is considered as a fundamental scale. A running of the gauge parameters $\lambda$ and $g$ could cause them to weigh the various tensors in the Euler-Lagrange equations differently, making some of them negligible compared to others. In other words, the dynamics we have described could emerge as effective dynamics resulting from the quantum nature of the theory. 
Finally, the meaning of the contraction parameter $\alpha$ and its interplay with the cosmological constant have to be understood, also in relation with the renormalization group flow. 

As a final speculation, we observe that, since the dynamical metric is quadratic in the sector of P-translations of the gauge algebra,  the undelying gauge theory could be a natural candidate for  a double copy formulation of gravity \cite{Campiglia:2021srh, Monteiro:2011pc,BCJ}.
 Focused work in this direction would require specifying solutions of the parent YM mapped to the gravitational sector. In this case, the interest of our group-theoretic reduction lies in the fact that the algebra characterizes the field solution whereas, in the double copy procedure, the gauge algebra of the YM field is generally unspecified.  In this sense, recent work \cite{LopesCardoso:2024ttc} investigates the relation between gravity and self-dual YM (SDYM) in 4D by dimensional reduction, via a shared description for specific sets of solutions of the two theories in terms of a 2D integrable Breitenlohner-Maison system and a \textit{single-copy} lift back to 4D SDYM. In this case, the Breitenlohner-Maison procedure itself determines the Yang-Mills gauge group, encoding aspects of the gravitational solution in the Yang-Mills gauge group generators supporting the solution. We plan to address the possible connection with the double copy scenario and in particular with the findings of \cite{LopesCardoso:2024ttc} in future work. 

\section*{Acknowledgments}
The authors wish to thank Leonardo Castellani for pointing out important issues in the first version of the paper and Stefano Liberati for interesting discussions. They acknowledge support from the INFN Iniziativa Specifica GeoSymQFT and from  the European COST Action CaLISTA CA21109.  P.V. acknowledges support from the  Programme STAR Plus, financed  by UniNA and Compagnia di San Paolo, and from  the PNRR MUR Project No. CN 00000013-ICSC.
\appendix

\section{THE (A)dS GROUPS}
\label{wtr}
In this appendix we fix the notation and our convention for the structure groups $SO({p},{q})$ of the Yang-Mills model analyzed in the paper.
These are the pseudo-orthogonal groups $SO({p},{q})$ with  $p+q=5$,  namely the special Lie groups of linear transformations of a real five-dimensional vector space that leave invariant a nondegenerate symmetric bilinear form $\eta$ of signature $(p,q)$.  
The algebra $\mathfrak{so}(p,q)$ is provided by generators $M_{AB}$,  $A,B=0,\dots, 4$, and  Lie brackets
\begin{equation}
[M_{AB}, M_{CD}] = \eta_{AD} M_{BC} - \eta_{AC} M_{BD} + \eta_{BC} M_{AD} - \eta_{BD} M_{AC}:=f^{KL}_{AB,CD}M_{KL}
\end{equation}
The Killing-Cartan metric for $\mathfrak{so}(p,q)$ is, by definition, 
\begin{equation}\label{killig-cartan-metric}
(M_{AB}, M_{CD})= f^{KL}_{AB,EF} f^{EF}_{CD,KL} = (2p + 2q - 4)(\eta_{AD}\eta_{BC} - \eta_{AC}\eta_{BD})
\end{equation}
Specifically we are interested in the de Sitter group $SO(1,4)$ preserving the quadratic form $\eta_{AB}=diag(1,-1,-1,-1,-1)$ and the anti-de Sitter group $SO(2,3)$ preserving  $\eta_{AB}=diag(1,-1,-1,-1,1)$. Our choice of basis for their Lie algebra is the one that singles out the Lorentz subalgebra. Indeed, if $a,b=0,1,2,3$ we can write
\begin{equation}
\begin{split}
    [M_{4a},M_{4b}]&=\pm M_{ab}
\\
[M_{4a},M_{bc}]&=\eta_{ab}M_{4c}-\eta_{ac}M_{4b}
\\
[M_{ab},M_{cd}]&=\eta_{ad} M_{bc} - \eta_{ac} M_{bd} + \eta_{bc} M_{ad} - \eta_{bd} M_{ac}
\end{split}
\end{equation}
where + and $-$ refer, respectively, to the de Sitter  and anti-de Sitter algebra. Hence we  define 
\begin{equation}
P_a=\sqrt{|\alpha|}M_{4a}
\end{equation}
\begin{equation}
J_{ab}=M_{ab}
\end{equation}
so that
\begin{equation}
\label{P-algebra}
\begin{split}
[P_a,P_b]&=\pm|\alpha|  J_{ab}:=\alpha K_{a,b}^{ef}J_{ef}
\\
[P_a,J_{bc}]&=\eta_{ab}P_c-\eta_{ac}P_b:=f^e_{a,bc}P_e
\\
[J_{ab},J_{cd}]&=\eta_{ad} J_{bc} - \eta_{ac} J_{bd} + \eta_{bc} J_{ad} - \eta_{bd} J_{ac}:=f^{ef}_{ab,cd}J_{ef}
\end{split}
\end{equation}
where 
$\eta_{ab}$ is the Minkowski metric in 4D with signature  $(+,-,-,-)$
while 
$K^{cd}_{a,b}=\frac{1}{2}(\delta^c_a\delta^d_b-\delta^c_b\delta^d_a)$. The parameter $\alpha$  can take positive, negative, or zero values. The Poincaré algebra $\mathfrak{iso}(1,3)$ can be derived from $\mathfrak{so}(1,4)$ or $\mathfrak{so}(2,3)$ in the limit of $\alpha\rightarrow 0$ (the $In\ddot{o} n\ddot{u}$-$Wigner$ contraction \cite{Inonu}). Finally, by Eq. (\ref{killig-cartan-metric}),  the scalar product of the generators of the algebra (\ref{P-algebra}) is given by
\begin{align}
\label{prodotto finale m e P}
 (P_{a}, P_{b}) &= |\alpha|(M_{4a}, M_{4b}) = 6|\alpha|(\eta_{4b}\eta_{a4}-\eta_{44} \eta_{ab})
 \\ \nonumber
(J_{ab}, P_{c})&=\sqrt{|\alpha|}(M_{ab}, M_{4c}) = 6\sqrt{|\alpha|}(\eta_{ac} \eta_{b4} - \eta_{bc} \eta_{a4}) = 0\\ \nonumber
(J_{ab}, J_{cd})&=(M_{ab}, M_{cd})=6(\eta_{ac} \eta_{bd} - \eta_{bc} \eta_{ad}) \nonumber
\end{align}
or, in compact form
\footnote{The numerical coefficient emerging in Eq (\ref{prodotto finale m e P}), can be reabsorbed by redefining the coupling constant of the action.} 
\begin{equation}
\label{Finale (J,J) e (JP)}
\begin{aligned}
    (P_a, P_b) &= \pm|\alpha|\eta_{ab}=\alpha\eta_{ab}, \\
    (J_{ab}, P_c) &= 0, \\
    (J_{ab}, J_{cd}) &= \eta_{ac}\eta_{bd} - \eta_{bc}\eta_{ad}
\end{aligned}
\end{equation}
which becomes degenerate in the limit where $\alpha$ is zero, i.e., in the limit where the algebra becomes $\mathfrak{iso}(1,3)$.

\bibliographystyle{utphys}
\bibliography{refs}
\end{document}